\documentclass{article}

\usepackage{latexsym}
\usepackage{amsmath,amssymb,amsthm,amsfonts,color,colordvi,url}
\usepackage[]{graphicx}
\usepackage{subfigure}

\newcommand{\ket}[1]{\vert#1\rangle}

\newcommand{\expect}[1]{\langle #1 \rangle}
\newcommand{\absolute}[1]{\vert #1 \vert}

\newtheorem{theo}{Theorem}
\newtheorem{defi}{Definition}
\theoremstyle{definition}
\newtheorem{lemma}{Lemma}
\newtheorem{corol}{Corollary}

\newcommand{\phiplus}{\frac{1}{\sqrt{2}}
\ket{000}+\frac{1}{\sqrt{2}}\ket{111}}

\newcommand{\dDJ}{distributed
Deutsch-Jozsa }
\def\floor#1{\lfloor #1 \rfloor}
\def\ceil#1{\lceil #1 \rceil}
\newcommand{\A}{^{(A)}}
\newcommand{\B}{^{(B)}}
\newcommand{\C}{^{(C)}}
\newcommand{\ii}{^{(i)}}
\newcommand{\jj}{^{(j)}}

\begin{document}

\title{On the power of non-local boxes}

\author{
Anne Broadbent \hspace{0.75cm} Andr\'e Allan M\'ethot\\[0.5cm]
\normalsize\sl D\'epartement d'informatique et de recherche op\'erationnelle\\[-0.1cm]
\normalsize\sl Universit\'e de Montr\'eal, C.P.~6128, Succ.\ Centre-Ville\\[-0.1cm]
\normalsize\sl Montr\'eal (QC), H3C 3J7~~\textsc{Canada}\\
{\normalsize\texttt{\{broadbea,\,methotan\}}\textbf{\char"40}\texttt
{iro.umontreal.ca}}}

\date{Revised August 19, 2005}
\maketitle


\begin{abstract}
A non-local box is a virtual device that has the following property:
given that Alice inputs a bit at her end of the device and that Bob
does likewise, it produces two bits, one at Alice's end and one at
Bob's end, such that the $XOR$ of the outputs is equal to the $AND$
of the inputs. This box, inspired from the CHSH inequality, was
first proposed by Popescu and Rohrlich to examine the question:
given that a maximally entangled pair of qubits is non-local, why is
it not maximally non-local? We believe that understanding the power
of this box will yield insight into the non-locality of quantum
mechanics. It was shown recently by Cerf, Gisin, Massar and Popescu,
that this imaginary device is able to simulate correlations from any
measurement on a singlet state. Here, we show that the non-local box
can in fact do much more: through the simulation of the magic square
pseudo-telepathy game and the Mermin-GHZ pseudo-telepathy game, we
show that the non-local box can simulate quantum correlations that
no entangled pair of qubits can, in a bipartite scenario and even in
a multi-party scenario. Finally we show that a single non-local box
cannot simulate all quantum correlations and propose a
generalization for a multi-party non-local box. In particular, we
show quantum correlations whose simulation requires an exponential
amount of non-local boxes, in the number of maximally entangled
qubit pairs.
\end{abstract}


\section{Introduction}\label{intro}

In a 1964 influential paper, Bell showed that there exist
correlations that can be obtained from bipartite measurements of a
quantum state that no local realistic theory can
reproduce~\cite{bell64}. From this, if one believes that quantum
mechanics is a correct description of the world, one is forced to
conclude that Nature is fundamentally non-local. This astounding
discovery has lead to a rich and still developing literature. One of
the best known papers in the field is the 1969 experimental
proposition of Clauser, Horne, Shimony and Holt~\cite{chsh69}. The
authors put forth an inequality which all local hidden variable
(LHV) models must satisfy:
\begin{equation}\label{classicalinequality}
\absolute{\expect{A_1 B_1} + \expect{A_1 B_2} + \expect{A_2 B_1} -
\expect{A_2 B_2}}\leq 2,
\end{equation}
where $A_1$ and $A_2$ are local spin measurements of a spin-half
particle on Alice's subsystem and $B_1$ and $B_2$ are measurements
on Bob's subsystem. While any LHV model has to abide by this rule,
quantum mechanics can violate Inequality~\ref{classicalinequality}
by an appropriate choice of measurements on a maximally entangled
state, such as $\ket{\psi^-}= (\ket{+-}-\ket{-+})/\sqrt{2}$:
\begin{equation}\label{quantuminequality}
\absolute{\expect{A_1 B_1} + \expect{A_1 B_2} + \expect{A_2 B_1}
-\expect{A_2 B_2}}= 2\sqrt{2}.
\end{equation}

This result may also be interpreted in a more intuitive
fashion~\cite{BCD}: if Alice and Bob want to play a game,
called the CHSH game, where they are each given as input a bit,
$x\A$ and $x\B$ respectively, and they want to produce output bits
$y\A$ and $y\B$ respectively such that
\begin{equation}\label{nlb}
    x\A \wedge x\B = y\A \oplus y\B,
\end{equation}
then there is no classical (LHV) strategy that can help them win
this game with a probability greater than $3/4$, but if they share
the quantum state $\ket{\psi^-}= (\ket{01} -\ket{10})/\sqrt{2}$,
then they can succeed with probability $\cos^2(\pi/8)\approx
0.85$~\cite{chsh69}.

Many years later, Popescu and Rohrlich~\cite{pr94} asked a natural
question: why not more? Given that quantum mechanics is non-local,
why is it not maximally non-local? Many authors have studied this
question~\cite{cirelson80,chtw04,vandam00,vandam05,bblmtu05} and
we will discuss their results in Section~\ref{spt}. Besides this
intriguing question, Popescu and Rohrlich suggested something else
of interest, a \emph{gedanken} product: the non-local box (NLB). A
NLB is a virtual device that has two ends and the following
property: if Alice inputs a bit into her end of the NLB and Bob
does likewise, then they will both receive  a bit from the NLB
such that the condition of Equation~\ref{nlb} is always respected,
and such that all solutions are equally likely. It is important to
note that this device does not allow faster than light
communication~\cite{pr94}.

Recently, Cerf, Gisin, Massar and Popescu built on the work of Toner
and Bacon~\cite{tb03} and used a NLB to simulate the correlations
obtained from any bipartite measurement of a maximally entangled
pair of qubits, $\ket{\psi^-}$, without any
communication~\cite{cgmp04}. This result shows that signaling
information on the inputs is not necessary for a perfect simulation
of quantum correlations. The long term aim of this work is to
characterize the NLB in order to yield insights into the
non-locality of Nature.

In this paper, we want to push this research further. The NLB was
inspired from the CHSH inequality, which is often thought as the
generic inequality for non-locality, and it can also simulate the
correlations of a maximally entangled pair of qubits.  From this,
it is tempting to draw an analogy between the NLB and the
maximally entangled pair of qubits. We will show however that a
single NLB can be used to accomplish a distributed task that
cannot be accomplished with only a maximally entangled pair of
qubits. In particular, we will study pseudo-telepathy and show
simulations of some pseudo-telepathy games with one NLB where the
quantum strategy requires more than a maximally entangled pair of
qubits to succeed. We will also give limitations on what a single
NLB can achieve and propose a generalization of the NLB to the
multi-party setting.

\begin{defi}
A \emph{bipartite game} $G=(X,Y,R)$ is a set of inputs $X=X\A\times
X\B$, a set of outputs \mbox{$Y=Y\A\times Y\B$} and a relation
$R\subseteq X\A\times X\B\times Y\A\times Y\B$.
\end{defi}

\begin{defi}
A \emph{winning strategy} for a bipartite game  $G=(X,Y,R)$ is a
strategy according to which for every   $x\A\in X\A$ and $x\B\in
X\B$, Alice and Bob output $y\A \in Y\A$ and $y\B \in Y\B$
respectively such that $(x\A,x\B,y\A,y\B)\in R$.
\end{defi}

\begin{defi}
We say that a bipartite game $G$ exhibits \emph{pseudo-telepathy}
if  bipartite measurements of an entangled quantum state can yield
a winning strategy, whereas no classical strategy that does not
involve communication is a winning strategy.
\end{defi}
The generalization to multi-party pseudo-telepathy to be taken is
the natural one. For a complete survey on pseudo-telepathy, please see~\cite{bbt04}.

\begin{defi}
A \emph{non-local} protocol is a purely classical protocol where
the participants are not allowed communication but are allowed the use of NLBs.
\end{defi}

\begin{defi}
A protocol \emph{simulates} the \emph{correlations} of a
pseudo-telepathy game if, in addition to yielding a winning
strategy, the probabilities $Pr(Y\A,Y\B | X\A, X\B)$ are identical
to those of a quantum winning strategy.
\end{defi}


\section{Magic square game}\label{msg}

We saw in Section~\ref{intro} that one use of an NLB can give the
correlations of any bipartite measurement on $\ket{\psi^-}$
without any communication. A natural question would be to ask
whether it can give us more. In particular, are there correlations
that can only be obtained by bipartite measurements of an
entangled state of more than a pair of qubits, but that can be
simulated with one use of an NLB? In this Section, we answer
affirmatively by showing a pseudo-telepathy game, the magic square
game~\cite{aravind02}, that requires more than an entangled state
of two qubits in the quantum winning strategy, yet only one use of
an NLB suffices to yield a non-local winning strategy.  We also
give a non-local strategy that makes use of a single NLB and  that
simulates the magic square correlations.

\begin{defi}\label{defmsg}
In the \emph{magic square game}, Alice and Bob are given
 $x\A \in \{1,2,3\}$ and $x\B \in \{1,2,3\}$, respectively. They
produce 3 bits each, $(y_1\A,y_2\A,y_3\A)$ and
$(y_1\B,y_2\B,y_3\B)$, such that:
\begin{equation}
\label{magicsquare}
\begin{split}
y_3\A &= y_1\A\oplus y_2\A \\
y_3\B &= y_1\B\oplus y_2\B\oplus 1 \\
y_{x\B}\A &= y_{x\A}\B.
\end{split}
\end{equation}
\end{defi}
Here, and in all future definitions of bipartite games, it is
understood that $(x\A, x\B, y\A, y\B) \in R $ if and only if the
given equations are satisfied.

It is known that the magic square game is a pseudo-telepathy game:
the best classical players can do is succeed on $8/9$ of the
possible inputs, whereas players with the shared entangled state
 $\ket{\psi}
= \frac{1}{2} \ket{0011} -\frac{1}{2}\ket{0110}-
\frac{1}{2}\ket{1001} + \frac{1}{2} \ket{1100}$ (two maximally
entangled pairs of qubits), where Alice has the first two qubits
and Bob the last two qubits, have a quantum winning
strategy~\cite{bbt04}.

It is useful here to mention that a \emph{magic square} is a $3
\times 3$ matrix with binary entries such that the sum of each row
is even and the sum of each column is odd. It is obvious that such a
magic square does not exist, yet Alice and Bob's output bits (as
defined in Equation \ref{magicsquare}) fit perfectly into a magic
square: we place Alice's three output bits in the $x\A$th row and
Bob's three output bits in the $x\B${th} column. Using this same
construction, we can represent a player's strategy as a $3\times 3$
binary matrix.

\begin{lemma}\label{quantumstratmsg}
No quantum strategy can win the magic square game with probability
one if the participants share only  an entangled pair of qubits,
$\ket{\psi}=\alpha\ket{00}+\beta\ket{11}$.
\end{lemma}
\begin{proof}
The proof is  straightforward from  Brassard, M\'ethot and
Tapp~\cite{bmt04}, where the authors show that there cannot exist
a protocol that exhibits pseudo-telepathy where the quantum
strategy makes use of a  pair of entangled qubits.
\end{proof}

\begin{theo}\label{commmsg}
The magic square game  can be won classically with probability one
if the participants are allowed one bit of communication.
\end{theo}
\begin{proof}
\begin{figure}[h]%
\centering

\subfigure[Alice]{\nonumber  \label{subfig1}
\begin{tabular}{|c|c|c|} \hline 0&1&1
\\ \hline 1&1&0
\\ \hline 0&1&1 \\ \hline
\end{tabular} }
\subfigure[Bob]{ \label{subfig2}
\begin{tabular}{|c|c|c|} \hline 0&1&1 \\ \hline 1&1&0
\\ \hline 0&1&0 \\ \hline
\end{tabular}}
\subfigure[Alice]{
 \label{subfig3}
\begin{tabular}{|c|c|c|}
\hline 0&1&1 \\ \hline 1&1&0 \\ \hline 0&1&1 \\ \hline
\end{tabular}}
\subfigure[Bob]{
 \label{subfig4}
\begin{tabular}{|c|c|c|}
\hline 0&1&1 \\ \hline 1&1&1 \\ \hline 0&1&1 \\ \hline
\end{tabular}}

\caption{Two strategies: strategy 0 (\subref{subfig1} and
\subref{subfig2})  and strategy 1  (\subref{subfig3} and
\subref{subfig4}).}
 \label{Strategies}
\end{figure}
Alice and Bob agree ahead of time on a two strategies, say 0 and 1.
Strategy 0  yields a correct answer for all inputs except when
$x\A=x\B=3$, and strategy 1 yields a correct answer when
$x\A=x\B=3$.  Furthermore, strategies 0 and 1 can be chosen such
that Alice's outcomes are identical for both strategies. We give an
example of such strategies in Figure \ref{Strategies}. Alice and
Bob's final strategy is for Alice to send a single bit to Bob,
indicating whether or not $x\A =3$.  If $x\A \neq 3$, Bob acts
according to strategy 0, otherwise he uses strategy 1. It is easy to
check that with this strategy, Alice and Bob always win.
\end{proof}

\begin{theo}\label{theorem:classsimmsg}
Classical players that are  allowed one bit of communication can
simulate the  magic square correlations.
\end{theo}

\begin{proof}
Since, in the quantum strategy, Alice's and Bob's density matrices
are totally mixed, the local outputs of their von Neumann
measurements are uniformly distributed among all possible outputs
respecting the conditions of Definition~\ref{defmsg}.

Now in the classical protocol, Alice and Bob agree on strategies
$0$ and $1$ as in the proof of Theorem~\ref{commmsg}, but they use
shared randomness to choose the strategies uniformly at random
among all strategies that fit the construction.  With this
strategy, Alice and Bob's outcomes are distributed uniformly at
random among all possible winning outcomes.\end{proof}

\begin{theo}\label{theorem:MSnon-local}
There exists a non-local winning strategy for the magic square
game that makes use of a single NLB.
\end{theo}
\begin{proof}
Alice and Bob each have two strategies, say $A0$ and $A1$ for Alice
and $B0$ and $B1$ for Bob. Both of Alice's strategies respect the
condition $y_3\A= y_1\A\oplus y_2\A$ and Bob's $y_3\B= y_1\B\oplus
y_2\B \oplus 1$. Both pairs of strategies $(A0,B0)$ and $(A1,B1)$
yield a correct answer, $y_{x\B}\A= y_{x\A}\B$, for all inputs
except when $x\A=x\B=3$. Additionally, strategies $A0$ and $B1$, as
well as $A1$ and $B0$, are coordinated such that if Alice answers
according to strategy $Ai$ ($i \in{0,1}$) and Bob according to
strategy $Bj$ ($j=i \oplus 1$), then on inputs $x\A=x\B=3$, we have
that $y_3\A = y_3\B$. Such strategies ($A0,A1,B0$ and $B1$) are easy
to find.

Alice and Bob use an NLB to determine which strategy each player
uses: they both input in the NLB whether $x\A =3 $ or whether $x\B =
3$.  They then independently use the output of the NLB, $z\A$ and
$z\B$ to determine the strategy to use ($A z\A$ for Alice, $B z\B$
for Bob).

Note that by virtue of the NLB, Alice and Bob will have $z\A=z\B$ as
long as $x_A \neq 3$ or $x_B \neq 3$. Strategies $(A0,B0)$ and
$(A1,B1)$  will yield correct answers in this case.  If, however,
both $x\A = 3$ and $x\B = 3$, then Alice and Bob will answer
according to strategies $(A0,B1)$ or $(A1,B0)$. But these strategies
are coordinated so that $y_3\A = y_3\B$, so their answer is correct.
\end{proof}

\begin{theo}\label{simmsg}
There exists a non-local protocol that simulates the magic square
correlations with a single use of an NLB.
\end{theo}

\begin{proof}
The proof is similar to the proof of Theorem
\ref{theorem:classsimmsg}: all that Alice and Bob must do in order
to simulate the magic square correlations is  apply the strategy
given in the proof of Theorem \ref{theorem:MSnon-local}, but with
strategies $A0$, $A1$, $B0$ and $B1$ chosen among all possible
such strategies according to the uniform distribution.  Then Alice
and Bob's outcomes are distributed uniformly at random and
Definition~\ref{defmsg} is satisfied.
\end{proof}

{From} Lemma~\ref{quantumstratmsg} and Theorem~\ref{simmsg}, we get
the following Corollary:

\begin{corol}
An NLB can simulate bipartite correlations that no entangled pair
of qubits, $\ket{\psi}=\alpha\ket{00}+\beta\ket{11}$, can.
\end{corol}


\section{Mermin--GHZ game}\label{merminghz}

In this Section, we  add to the demonstration of the power of a
NLB by showing that it can also simulate correlations found in a
tripartite state.

\begin{defi}\label{defmermin}
In the \emph{Mermin--GHZ} game~\cite{mermin90a}, Alice, Bob and
Charlie are each given a bit such that \mbox{$x\A + x\B + x\C
\equiv 0 \pmod 2$} and they must produce a bit of output each,
$y\A$, $y\B$ and $y\C$, such that:
\[
 y\A \oplus y\B \oplus y\C = \frac{x\A + x\B + x\C}{2}.
\]
\end{defi}

It is well known that this is a pseudo-telepathy game.  In the
quantum winning strategy, Alice, Bob and Charlie share a
\emph{GHZ-state}: $\smash{\phiplus}$.

\begin{lemma}\label{quantumstratmermin}
No quantum strategy can win the Mermin--GHZ game with probability
one if any two participants share only an entangled pair of
qubits, $\ket{\psi}=\alpha\ket{00}+\beta\ket{11}$.
\end{lemma}

\begin{proof}
As in the proof of Lemma \ref{quantumstratmsg}, the result follows
from \cite{bmt04}. \end{proof}

\begin{theo}\label{commmermin}
The Mermin--GHZ game can be won classically with probability one if
the participants are allowed one bit of communication.
\end{theo}

\begin{proof}
The classical strategy that uses a bit of communication is the
following:  Bob and Charlie output $y\B =b$, $y\C=c$ respectively
where $b$ and $c$ are arbitrary bits known to all participants. Bob
sends $x\B$ to Alice, who computes $y=x\A \vee x\B$ and outputs
$y\A= b \oplus c \oplus y$.  It is easy to check that this strategy
works.
\end{proof}

\begin{theo}\label{commmerminSim}
The Mermin--GHZ correlations can be simulated by classical
participants using a single bit of communication.
\end{theo}

\begin{proof}
First, note that the quantum winning strategy (as given in
\cite{bbt04}, for instance) is such that the outcomes of the
players are uniformly distributed among all outcomes satisfying
 Definition~\ref{defmermin}. Now, Alice and Bob can used shared
randomness to select uniformly at random among all strategies that
succeed in the proof of Theorem \ref{commmermin}. This gives a
simulation of the Mermin--GHZ correlations.
\end{proof}

\begin{theo}\label{thm:merminNLB}
The Mermin--GHZ game can be won with probability one if the
participants are allowed one use of an NLB.
\end{theo}

\begin{proof}
Once again, we will use the NLB in our construction to replace the
communication in the protocol of Theorem~\ref{commmermin}. First,
we note the relationship between the logical $OR$ and the logical
$AND$:
\[
x\A \vee x\B = \overline{\bar{x}\A \wedge \bar{x}\B}.
\]
The strategy is then simple. Alice and Bob flip their inputs and
feed them into a shared  NLB which returns $y\A$ and $y\B$ such
that
\[
y\A \oplus y\B = \overline{x\A \vee x\B}.
\]
Since $x\A + x\B + x\C \equiv 0 \pmod 2$,
\begin{equation*}
\overline{x\A \vee x\B}=\left( \frac{x\A + x\B + x\C}{2} \right)
\oplus 1.
\end{equation*}
If Charlie outputs $y\C=1$, the protocol satisfies
Definition~\ref{defmermin}.
\end{proof}

\begin{theo}\label{simmermin}
There is a non-local protocol that simulates the Mermin--GHZ
correlations with a single use of an NLB.
\end{theo}

\begin{proof} As in the proof of Theorem~\ref{commmerminSim},
we can randomize the proof of Theorem \ref{thm:merminNLB}
so that the outcomes of Alice, Bob and Charlie are uniformly
distributed among all outcomes that satisfy
Definition~\ref{defmermin}. All we need to add is a random bit
shared  between the participants telling whether or not Bob and
Charlie should both flip their outputs or not.
\end{proof}

{From} Lemma~\ref{quantumstratmermin} and Theorem~\ref{simmermin},
we get the following Corollary:

\begin{corol}
An NLB can simulate tripartite correlations that no entangled pair
of qubits, $\ket{\psi}=\alpha\ket{00}+\beta\ket{11}$, can.
\end{corol}


\section{Non-local box pseudo-telepathy}\label{spt}

We have seen in Sections~\ref{msg} and~\ref{merminghz} that a
single use of an NLB can simulate quantum correlations that are
stronger than those obtained by bipartite measurements of a
maximally entangled pair of qubits. Can an NLB  do more? In this
Section, we discuss the known result that an NLB can indeed yield
correlations that cannot be reproduced by quantum mechanics by
showing an NLB pseudo-telepathy game that can be won with
probability one with a single use of an NLB while no quantum
protocol can.

\begin{defi}\label{defspt}
We say that a bipartite game exhibits \emph{non-local box
pseudo-telepathy} if there exists a non-local winning strategy,
while no winning strategy based on the laws of quantum mechanics
exists.
\end{defi}

\begin{lemma}\label{sCHSH}
A single NLB is sufficient to yield a protocol for an NLB
pseudo-telepathy game.
\end{lemma}

The game in which we are interested  is what the NLB is defined to
do. It is clear from the definition of the NLB that, using a such
a device, Alice and Bob can produce outputs such that the $XOR$ of
their outputs is equal to the $AND$ of their inputs. When Popescu
and Rohrlich proposed the NLB, it was already known, although not
expressed in these terms, that it could yield NLB
pseudo-telepathy.

In fact, in 1980, Tsirelson~\cite{cirelson80} showed that quantum
mechanics could not yield a value greater than $2\sqrt{2}$ in
Equation~\ref{quantuminequality} while, by definition, the NLB has
the algebraic maximum value of 4. Cleve, H\o yer, Toner and
Watrous~\cite{chtw04} generalized Tsirelson's result to show that
there cannot be a bipartite game with binary outputs that cannot be
won classically with probability one while a quantum protocol could.
Since the CHSH game cannot be won classically with probability
greater than $3/4$, then no quantum strategy can win with
probability~1. More recently, van Dam~\cite{vandam00,vandam05} and
others~\cite{bblmtu05}, also showed that no quantum strategy can win
the CHSH game with probability equal to unity by taking an
altogether different approach. They showed how we can use
NLBs~\cite{vandam00,vandam05}, or even faulty NLBs~\cite{bblmtu05},
to reduce all of communication complexity for decision problems to a
single bit. Since we know that quantum communication complexity is
not trivial~\cite{cdnt97}, no quantum simulation of the NLB can
exist.


\section{Limits on the power of the non-local box}\label{limits}

In previous Sections, we have shown the amazing power of a single
NLB. We have demonstrated quantum correlations that cannot be
generated by an entangled pair of qubits but still can be
simulated with only one NLB. Do all quantum correlations collapse
to a single use of an NLB? The answer is no. In \cite{bgs05}, it is
shown that one use of an NLB is not sufficient to simulate
non-maximally entangled states of two qubits. Here, we will also
prove that there exist pseudo-telepathic correlations (whose
simulation cannot require more resources than the simulation of
general measurements on the quantum state used in the quantum
winning strategy) that cannot be simulated with a single NLB. We
will first show that in a multi-party setting, there exist
pseudo-telepathic correlation that require more than one use of a
NLB to simulate. We then use the distributed Deutsch-Jozsa game to
show that some bipartite pseudo-telepathic correlations also
require more than one use of an NLB to simulate. As a consequence,
we will prove that maximally entangled bipartite states and NLBs
are truly different resources.

\begin{defi}\label{mmermin}
The \emph{multi-party Mermin--GHZ} game \cite{Mermin, Recasting} is defined as
follows. Each player $i \in \{1, \ldots , n\}$ $(n \geq 3)$ is given
a bit $x\ii$ such that $\sum_i x\ii \equiv 0 \pmod 2$. Each player
must produce a bit $y\ii$ of output such that:
\[
\sum_i y\ii \equiv \left(\frac{\sum_i x\ii}{2}\right) \pmod 2.
\]
\end{defi}

\begin{theo} \label{thm:sufficient}
$\binom{n}{2} \in O(n^2)$ NLBs are sufficient for the simulation
of the multi-party Mermin--GHZ correlations.
\end{theo}

\begin{proof}
Each player shares an NLB with every other player (there are
therefore $\binom{n}{2} $ NLBs).  Upon receiving his input $x\ii$,
player $i$ feeds $x\ii$  into each of his  shared NLBs. Let
$y^{(i,j)}$ be the output of the NLB shared with player $j$.
Player $i$ then computes the parity of all such $y^{(i,j)}$: let
$y\ii=\sum_{j\neq i} y^{(i,j)} \pmod 2$.  This is player $i$'s
output.

To show that this strategy works, note that
\begin{equation*}
\sum_i y\ii \equiv \sum_i\sum_{j\neq i} y^{(i,j)} \pmod 2,
\end{equation*}
and furthermore, $\forall i,j$ where $i\neq j$
 \begin{equation*}
y^{(i,j)} +  y^{(j,i)} \pmod 2 \equiv
\begin{cases}
0,&  x\ii \wedge x\jj =0  \\
1,&  x\ii \wedge x\jj =1\\
\end{cases}.
\end{equation*}
Therefore,  if $\sum_i x\ii = 4k$ for some non-negative integer
$k$, (and so $\left(\frac{\sum_i x\ii}{2}\right) \equiv 0 \pmod
2$), then \mbox{$\sum_i y\ii \equiv \binom{4k}{2} \equiv 0 \pmod
2$}. And if $\sum_i x\ii = 4k +2$ for some non-negative integer
$k$, (and therefore, $\left(\frac{\sum_i x\ii}{2}\right) \equiv 1
\pmod 2$), then $\sum_i y\ii \equiv \binom{4k+2}{2} \equiv 1 \pmod
2$.
\end{proof}

\begin{theo}\label{thm:necessary}
Any simulation of the multi-party Mermin--GHZ correlations for $n
\geq 4$ players requires more than a single use of an NLB.
\end{theo}
\begin{proof}
Consider the case where $n=4$.  Without loss of generality,
suppose that players~1 and~2 share an NLB.  Let us assume
furthermore that players~1 and~2 are allowed unlimited
communication with each other.  We will show that even under this
stronger assumption, there is no winning strategy for the
multi-party Mermin--GHZ game.  It follows that the four players
cannot simulate the multi-party Mermin--GHZ correlations with a
single NLB.

Let us consider a subset of the possible inputs: $I=\{ (0,0,0,0),
(0,0,1,1), (0,1,0,1), (0,1,1,0) \}$. If we consider players~$1$ and~$2$ as a single entity, we get, after relabelling, a new set of
inputs: $\{(0,0,0), (0,1,1), (1,0,1), (1,1,0)\}$. This is the
Mermin--GHZ game (Definition~\ref{defmermin}). Since a winning
strategy for the set $I$ of inputs leads
to a classical
winning strategy for the Mermin--GHZ game, which is impossible, this
contradiction proves our claim.

The result extends easily to the case of $n > 4$: even if we allow
communication between the first $n-2$ players,  we can find a
subset of inputs (as above) where the players need to be able to
win the Mermin--GHZ game in order to win this game.
\end{proof}

\begin{theo}
$\Omega(n)$ NLBs are necessary in a non-local winning strategy for
the multi-party Mermin--GHZ game.
\end{theo}
\begin{proof}
As we saw in the proof of Theorem~\ref{thm:necessary}, there
cannot be two players, or more, that are not linked with at least
one other player through an NLB. So in order for at least $n-1$
players to be linked with another player, we need
$\floor{n/2-1}+1\in \Omega(n)$ NLBs.
\end{proof}

We now turn to a bipartite scenario and show that there exist
bipartite quantum correlations that require more than one use of a
NLB to simulate.

\begin{defi}\label{defbct}
In the \emph{\dDJ}game~\cite{bct99}, Alice and Bob are given
$2^n$-bit strings $x\A$ and $x\B$ respectively such that
\begin{equation}\label{ddj}
\Delta(x\A , x\B) \in \{0,2^{n-1}\}
\end{equation}
where $\Delta(x\A , x\B)$ is the \emph{Hamming distance} between two
strings (Equation \ref{ddj} states that either the two strings are
the same or they differ in exactly half the bit positions). Then the
players  must output $n$-bit strings $y\A$ and $y\B$,
respectively such that:
\begin{equation}\label{eqn:DJ}
[y\A = y\B] \Leftrightarrow [x\A= x\B].
\end{equation}
\end{defi}

We know that for all $n \geq 4$, the above game is a
pseudo-telepathy game \cite{New04}, and the quantum state used for
the quantum winning strategy is \mbox{$\frac{1}{\sqrt{2^n}}
\sum_{j=0}^{2^n-1} \ket{j}\ket{j}$} \cite{bct99}. Furthermore, we
have the following lemma from~\cite{bct99}:

\begin{lemma}\label{commbct} A classical winning strategy for the \dDJ
game requires $\Omega(2^n)$ bits of communication.
\end{lemma}

\begin{theo}\label{simbct}
No classical winning strategy for the \dDJ game with less than $\Omega(2^n)$
uses of an NLB exists.
\end{theo}
\begin{proof}
Suppose we had a winning strategy for the  \dDJ game with less than
$\Omega(2^n)$ NLBs. Since we can simulate an NLB with one bit of
communication \cite{ww04}, we could use communication to transform
the winning strategy that uses NLBs into a winning strategy with
less than $\Omega(2^n)$ bits of communication (and no NLBs). Such a
strategy would contradict Lemma~\ref{commbct}.
\end{proof}

When considered as a resource, entanglement is usually quantified by
the number of maximally entangled bipartite states of two qubits,
$(\ket{00}+\ket{11})/\sqrt{2}$. In~\cite{bgs05}, Brunner, Gisin and
Scarani showed that there exist bipartite entangled states of two
qubits that \emph{cannot} be simulated with a single use of an NLB.
Since a single use of an NLB can simulate a maximally entangled
bipartite state of two qubits \cite{cgmp04}, the authors conclude
that ``entanglement and non-locality are different resources''. We
concur that according to their measure there is an anomaly which
also occurs in many other measures of non-locality~\cite{bgs05}.
However, when
 concerned with how many
 resources we need to perform a certain computational task,
we quantify resources in an asymptotic fashion. The result of~\cite{bgs05}
 is \emph{not} asymptotic: it does
not rule out a world in which $cn$ NLBs, for some constant $c$, are
sufficient to simulate $n$ bipartite entangled states. In such a
world, NLBs would still be considered strictly stronger than
entanglement, for when speaking of computational resources,
multiplicative constants do not matter. Our results have the
advantage of proving an asymptotic gap between the two resources: we
have shown that there exist correlations whose simulation requires
an exponential amount of NLB uses (in the number of maximally
entangled two qubit bipartite states). Furthermore, the existence of
NLB pseudo-telepathy games confirms that non-locality and
entanglement are different and incomparable resources.

Our result shows that the simulation of $n$ pairs of maximally
entangled qubits requires $\Omega(2^n)$ NLB uses. At first sight,
this may seem to contradict the fact that a single NLB use is
sufficient for the simulation of a single pair of maximally
entangled qubits. This apparent contradiction is explained by the
fact that, thanks to entanglement, the simulation of  $n$ bipartite
maximally entangled qubit pairs cannot, in general, be expressed as
$n$ independent simulations of separate systems of two qubits.

We finish this section by showing that the lower bound of Theorem
\ref{simbct} is tight.

\begin{theo}\label{djsimnlb}
There is a non-local winning strategy for the \dDJ game with
$O(2^n)$  NLB~uses.
\end{theo}

Before turning to the proof, first note that if the task were for
the players to outputs \emph{any} string $y\A$ and $y\B$
respectively, such that $[y\A = y\B] \Leftrightarrow [x\A= x\B]$,
then Alice and Bob could simply use $x\A$ and $x\B$ as outputs and
the condition is satisfied. The difficulty for Alice and Bob in
the \dDJ game is to output strings that are \emph{exponentially}
shorter than their inputs. In the following non-local winning
strategy, Alice and Bob will use NLBs to achieve this shorter
input.

Second, note that if Alice and Bob have two bits, $a_1, a_2$ and
$b_1, b_2$ respectively, then, making use of two NLBs, they can
compute bits $a$ for Alice and $b$ for Bob such that  $a \oplus b =
f(a_1, a_2, b_1, b_2)= (a_1 \oplus b_1) \wedge (a_2 \oplus b_2)$.
This observation follows from the fact that $f(a_1, a_2, b_1, b_2) =
a_1a_2 \oplus b_1b_2 \oplus a_1b_2 \oplus a_2b_1$, where the first
two terms can be computed locally, while the last two require one
use of an NLB each; Alice computes $A_1=a_1a_2$ and Bob $B_1=b_1b_2$,
Alice inputs $a_1$ into a first NLB while Bob inputs $b_2$, they get
$A_2$ and $B_2$ respectively and Alice inputs $a_2$ into a second
NLB while Bob inputs $b_1$ from which they get $A_3$ and $B_3$. With
$a=A_1\oplus A_2\oplus A_3$ and $b=B_1\oplus B_2\oplus B_3$, we
clearly have $a \oplus b= (a_1 \oplus b_1) \wedge (a_2 \oplus b_2)$.
We call such operation the \emph{distributed} computation of the
function $f$, which is analogous to computing the $AND$ of two
distributed bits, $a_1\oplus b_1$ and $a_2\oplus b_2$.\footnote{The
idea of using NLBs to replace communication in distributed
computations is due Cleve \cite{Cleve96} and van Dam
\cite{vandam00,vandam05}, who independently demonstrated that their
use allows any distributed Boolean function to be evaluated using a
single bit of communication.}

\begin{proof}
First, Alice flips all her input bits.  We'll call the resulting
string $\bar{x}\A$.  Using this new input, Alice and Bob execute a
series of \emph{rounds}. Each round $i$ has the following
property: at the beginning of the round, Alice has the string
$a^{(i)} \in \{0,1\}^{2^{n-i}}$ and Bob $b^{(i)} \in
\{0,1\}^{2^{n-i}}$ such that either the \emph{diametric}
($\Delta(a^{(i)},b^{(i)}) = 2^{n-i}$) or the \emph{disparity}
($\Delta(a^{(i)},b^{(i)}) < 2^{n-i}$) condition holds. At the end
of the round, Alice has the string $a^{(i+1)} \in
\{0,1\}^{2^{n-i-1}}$ and Bob $b^{(i+1)} \in \{0,1\}^{2^{n-i-1}}$
and the condition, diametric or disparity, is unchanged.

To execute round $i$, the players perform a sequence of
$2^{n-i-1}$ distributed computations of the function~$f$: for each
integer $j \in \{0, \ldots 2^{n-i-1} \}$, let $a_j^{(i+1)}$ and
$b_j^{(i+1)}$ be the result of the distributed computation of
$f(a_{2j}^{(i)}, a_{2j+1}^{(i)}, b_{2j}^{(i)}, b_{2j+1}^{(i)})$.
The final strings  for Alice and Bob at the end of round $i$ are
$a^{(i+1)}$ and  $b^{(i+1)}$, respectively.

It is easy to see that by virtue of the function $f$, if the
diametric condition holds at the beginning of the round, then it
still holds at the end of the round; the same is true for the
disparity condition.

Alice and Bob start round $0$ each with a $2^n$-bit string, $a^{(0)}
= \bar{x}\A$ and  $b^{(0)} = x\B$. They repeat many rounds until
they each have an $n$-bit string (they can pad their outputs with
diametric bit strings after the last round if necessary), therefore
performing $n-\floor{\lg n}$ rounds, for a total of $2 (
\sum_{i=0}^{n- \floor{\lg n}-1}2^{n-i-1}) = 2^{n+1} -2^{\floor{\lg n
}+1}  \in O(2^{n})$ NLBs. At the end of the sequence of rounds,
Alice flips the bits that she has calculated. The resulting strings
are $y\A$ for Alice and $y\B$ for Bob and from the diametric or
disparity condition, it is easy to see that $[y\A = y\B]
\Leftrightarrow [x\A= x\B]$.
\end{proof}


\section{A new game}\label{smerminghz}

We now attempt to answer the question: what is the generalization of
the NLB to a multi-party scenario? In \cite{cgmp04}, it is shown
that a natural extension of the NLB allows for instantaneous
signaling.  Here, we give a different extension: we give a new NLB
pseudo-telepathy game and propose a generalization of the NLB based
on this new game.

\begin{defi}\label{defgnlb}In this game, participant $i \in
\{1,\ldots n\}$ $(n \geq 2)$ is given a bit of input, $x\ii$.
 The participants must each output a bit  $y\ii$ such that:
\[
\sum\limits_{i=1}^{n} y\ii \pmod 2 = BMAJ(x^{(1)}, x^{(2)}, \dots
,x^{(n)})=
\begin{cases}
1\ \text{if }\Delta(x^{(1)}\, x^{(2)}\,\dots\, x^{(n)})>\lfloor n/2 \rfloor\\
0\ \text{otherwise}
\end{cases}
\]
where $BMAJ$ is simply the majority biased towards 0, and
$\Delta(x^{(1)}\, x^{(2)}\,\dots\, x^{(n)})$ is the \emph{Hamming
weight} of a bit string.
\end{defi}

\begin{theo} \label{thm:mpnlb}
There is no classical winning strategy for the game of
Definition~\ref{defgnlb}.
\end{theo}
\begin{proof}
For the case where $n=2$, this is exactly the task that an NLB
accomplishes.  We know that no classical strategy can succeed with
probability~1.  Now, for $n \geq 3$, we pick a subset $S$ of
possible inputs for which, even allowing communication between all
but two players yields a situation where no classical strategy can
succeed with probability~1: $S$ is the set of questions where the
first $\floor{\frac{n-2}{2}}$ players have input~$0$, the next
$\ceil{\frac{n-2}{2}}$ players have input~$1$ and the remaining
two players have inputs $0$ or~$1$.  Note that even by allowing
all players except the last two to communicate, we still get that
no classical strategy can succeed at this game, for a strategy to
win this game entails the existence of a strategy to win the CHSH
game described in Section~\ref{spt}.
\end{proof}

\begin{theo}\label{gnlbspt}
There is no quantum winning strategy for the game of
Definition~\ref{defgnlb}.
\end{theo}
\begin{proof}
For the case where $n=2$, this is exactly the task that an NLB
accomplishes.  We know that no quantum strategy can succeed with
probability~1.  Now, for $n \geq 3$, as in the proof of Theorem
\ref{thm:mpnlb}, we pick subset $S$ of possible inputs for which,
even allowing communication between all but two players yields a
situation where no quantum strategy can succeed with probability~1.
\end{proof}

\begin{theo}
$\Omega(n)$ NLBs are necessary in a non-local winning strategy for
the game of Definition~\ref{defgnlb}.
\end{theo}
\begin{proof}
As we saw in the proof of Theorem \ref{gnlbspt}, there cannot be
two players, or more, that are not linked with at least one other
player through an NLB. So in order for at least $n-1$ players to be
linked with another player, we need $\floor{n/2-1}+1\in \Omega(n)$
NLBs.
\end{proof}

\begin{theo}\label{sspt}
There is a non-local winning strategy for the game given in
Definition~\ref{defgnlb} with  $O(n^3 2^n)$ NLB~uses.
\end{theo}

The following scenario
is relevant to the proof of Theorem \ref{sspt}; it is a
generalization of the distributed computation of the function $f$
that we presented in the proof of Theorem~\ref{djsimnlb}.
 Consider $n$
participants. A bit $x_k$ is a called a \emph{distributed bit} if
each participant $i$ has a bit $x_k^{(i)}$ such that $ x_k =
\bigoplus_{i=1}^n x_k^{(i)}$.  We will see how we can compute a
distributed Boolean function on distributed bits with the help of
NLBs. First of all, if any player $i$ has a bit $x^{(i)}$, then a
distributed bit $x_k$ can be \emph{initialized} to the value
$x^{(i)}$ by letting $x_k^{(i)} = x^{(i)}$ and $x_k^{(j)} = 0$ for
all $j \neq i$. Next, it easy to see that the negation of a
distributed bit, say $\bar{x}_k$ can be computed  by requiring that
a single player flip his bit. Finally, the distributed $AND$ of two
distributed bits, $x_k$ and $x_\ell$, can be computed using NLBs
thanks to the following observation:
\begin{equation}\label{distAND}
\begin{split}
x_k\wedge x_\ell = & (x_k^{(1)}\oplus x_k^{(2)}\oplus \dots \oplus
x_k^{(n)})\wedge (x_\ell^{(1)}\oplus x_\ell^{(2)}\oplus \dots
\oplus x_\ell^{(n)})\\
= & x_k^{(1)}\wedge x_\ell^{(1)} \oplus x_k^{(2)}\wedge
x_\ell^{(2)} \oplus \dots  \oplus x_k^{(n)}\wedge
x_\ell^{(n)} \oplus \\
&  x_k^{(1)}\wedge x_\ell^{(2)} \oplus x_k^{(1)}\wedge
x_\ell^{(3)} \oplus \ldots \oplus  x_k^{(1)}\wedge x_\ell^{(n)}
\oplus
 \ldots
\oplus x_k^{(n)}\wedge x_\ell^{(n-1)}
\end{split}
\end{equation}
To calculate the distributed $x_m = x_k\wedge x_\ell$, each
participant performs a certain number of calculations, each
 yielding a single bit.  Each participant's final bit,
$x_m^{(i)}$ is the parity of the sum of all his calculated bits.
Now, the $n$ conjunctions  on the second-to-last row  of Equation
\ref{distAND} can be computed locally by each participant  and
each of the $n(n-1)$ conjunctions in the last row can be computed
with a single NLB. This shows how to calculate the  distributed
$x_k\wedge x_\ell$.
 We are now
ready to turn to the proof of Theorem \ref{sspt}.

\begin{proof}
To compute the distributed $BMAJ$, the players simply need to
output bits where the total parity of their output satisfies:
\begin{equation} \label{eqn:BMAJ}
\begin{split}
\sum_i y^{(i)} \!\! \pmod 2 = & (x^{(1)} \wedge x^{(2)} \wedge
\dots \wedge x^{(\lfloor n/2 \rfloor +1)}) \vee (x^{(1)} \wedge
x^{(3)}
\wedge \dots \wedge x^{(\lfloor n/2 \rfloor +2)}) \vee \dots \\
& \vee (x^{(\lfloor n/2\rfloor )} \wedge x^{(\lfloor n/2\rfloor +
1)} \wedge \dots \wedge x^{(n)}).
\end{split}
\end{equation}
The above Boolean formula comes from the simple observation that
$BMAJ = 1$ if and only if there is a  $\lfloor n/2 \rfloor
+1$-subset of  $\{x^{(1)}, x^{(2)}, \ldots ,x^{(n)} \}$, with each
element in the subset having value 1.  In Equation \ref{eqn:BMAJ},
we consider all such $\binom{n}{\lfloor n/2 \rfloor +1}$ possible
subsets. Furthermore, Equation \ref{eqn:BMAJ} can be translated into
a series of negations and $AND$ gates (using de Morgan's Law). We
wish to calculate the total number of $AND$ gates:  we have $\lfloor
n/2\rfloor$ $AND$ gates for each of the $\binom{n}{\lfloor n/2
\rfloor +1}$ conjunctions as well as $\binom{n}{\lfloor n/2 \rfloor
+1} -1$ $AND$ gates for the disjunctions (since an $OR$ gate can be
computed with a single $AND$ gate and negations).  The total number
 of $AND$ gates is therefore  $(\lfloor n/2\rfloor){\binom{n}{\lfloor n/2\rfloor
+1}}+{\binom{n}{\lfloor n/2\rfloor +1}}-1 \in O(n2^n)$.

To evaluate Equation \ref{eqn:BMAJ} in a distributed way, the
participants simply initialize a sequence of distributed bits and
perform a sequence of distributed $AND$ calculations (as described
above the present proof and according to Equation \ref{eqn:BMAJ}) .
Since our protocol computes $O(n2^n)$ distributed  $AND$s, using
$O(n^2)$ NLBs each, the protocol uses a total of $O(n^3 2^n)$ NLBs.
\end{proof}

We think that this new game should be taken to be the
generalization of the NLB to a multi-party NLB. The reasons are
multiple.

\begin{enumerate}
    \item This generalization yields exactly the NLB in a bipartite scenario.
    \item In the tripartite scenario, this new NLB simulates directly the Mermin--GHZ game
    \item It does not allow faster than light communication.
    \item The box is simple and elegant.
    \item We have shown in Theorem~\ref{gnlbspt} that this multi-party NLB exhibits NLB pseudo-telepathy for every $n \geq 2$.
    \item We think that this multi-party NLB
   exhibits correlations that require a large amount of bipartite NLB uses to simulate.
\end{enumerate}


\section{Conclusions}

In the present text, we have made progress towards characterizing
the remarkable power of the NLB. A single NLB can simulate
correlations that no entangled pair of qubits can: in the bipartite
scenario (Theorem~\ref{simmsg}), and in the multi-party scenario
(Theorem~\ref{simmermin}). In Section~\ref{spt}, we also showed that
the NLB can exhibit correlations that cannot be reproduced by
quantum mechanics and defined NLB pseudo-telepathy
(Definition~\ref{defspt}). Finally we showed in
Theorems~\ref{thm:necessary} and~\ref{simbct} that a single NLB
cannot reproduce all correlations of quantum mechanics and we
proposed in Definition~\ref{defgnlb} a generalization of the NLB to
the multi-party scenario which has a lot of desirable properties. By
showing that the simulation of some quantum correlations requires an
\emph{exponential} amount of NLBs in the number of shared entangled
qubit pairs (see Theorem~\ref{simbct}), and from the fact that NLB
pseudo-telepathy exists, we have demonstrated that NLBs and
entanglement are different, incomparable resources. The fact that
there are correlations that can be generated from NLBs and that
cannot come from any entangled state (see Sections~\ref{spt}
and~\ref{smerminghz}) further supports this conclusion. A single NLB
can generate correlations that are stronger than those that can be
provided by quantum mechanics and yet we still require an
exponential amount of NLBs for the simulation of certain quantum
correlations; in our opinion, this is due to the fact that  NLBs are
inherently classical and, as such, cannot be entangled with one
another.

The very attentive reader might have noticed a connection between
Theorem~\ref{commmsg} and Theorem~\ref{simmsg}, between
Theorem~\ref{commmermin} and Theorem~\ref{simmermin}, and between
Lemma~\ref{commbct} and Theorem~\ref{simbct}: we have transformed
classical strategies with $n$ bits of communication into protocols
with $n$ uses of an NLB. Can we always make this substitution? It is
of course not the case, for example in communication complexity, but
if we just want to simulate quantum correlations, signaling might
not be necessary. After all, entanglement alone cannot be used to
signal. A partial answer can be found in~\cite{bgs05}, in which the
authors proved that there exist correlations that can be generated
from a single bit of communication, constrained to not signal
information on the input, which cannot be simulated with an NLB. Even
though we cannot have a one-to-one equivalence, can the NLB
paradigm, without consideration to the number of NLBs, replace
communication that does not signal? The answer might not be easy to
find. Degorre, Laplante and Roland have recently built on the work
of M\'ethot~\cite{methot04} and Cerf, Gisin, Massar and
Popescu~\cite{cgmp04} to create a simulation of a maximally
entangled pair of qubits for any POVM using on average 2 NLBs and 4
bits of communication~\cite{julien}. In this construction, it might
not be easy to get rid of the communication since every simulation
of quantum entanglement  known to the authors that takes POVMs into
account is founded on a \emph{test}
principle~\cite{methot04,cgm99,mbcc00}: Bob receives some
information from Alice and tells her if it is satisfactory with what
he has, if not they start over. In order for Alice to know when to
start over, Bob must signal so to Alice. It is not clear if or how
we can get out of this test paradigm.

Of course, simulations of other pseudo-telepathy games need to be
done before we can claim to understand fully the NLB. In particular,
an open question of interest, and in relation to the discussion in
the previous paragraph, is whether any pseudo-telepathy game can be
simulated with NLBs. We would also like to see a non-trivial lower
bound for the number of NLBs required to simulate the generalization
to the multi-party setting put forward here and for the multi-party
Mermin--GHZ game.


\section*{Acknowledgements}

We would like to thank Julien Degorre, Patrick Hayden and J\"urg
Wullschleger for very stimulating discussions, and Gilles Brassard
and Alain Tapp for truly helpful comments. This work is supported in
part by Canada's {\sc Nserc}.


\end{document}